\pdfoutput=1

\documentclass[11pt]{article}

\usepackage[final]{acl}

\usepackage{times}
\usepackage{latexsym}

\usepackage[T1]{fontenc}

\usepackage[utf8]{inputenc}

\usepackage{microtype}

\usepackage{inconsolata}

\usepackage{graphicx}

\usepackage{booktabs}
\usepackage{amsmath}
\usepackage{array}
\usepackage{color}
\usepackage{bbding} 
\usepackage{amsfonts} 
\usepackage{algorithm}
\usepackage{algpseudocode}
\usepackage{multirow}
\usepackage{stfloats}
\usepackage[none]{hyphenat}

\definecolor{cred}{HTML}{FF6B6B}
\definecolor{cyellow}{HTML}{FEC260}
\definecolor{cgreen}{HTML}{70AD47}
\definecolor{cblue}{HTML}{4D96FF}
\definecolor{cpurple}{HTML}{2A0944}
\definecolor{ggray}{RGB}{127,127,127}
\definecolor{aliceblue}{rgb}{0.94, 0.97, 1.0}
\definecolor{cvprblue}{rgb}{0.21,0.49,0.74}

\title{Uni-Retrieval: A Multi-Style Retrieval Framework for STEM's Education}

\author{Yanhao Jia \textsuperscript{1${*}$},
        Xinyi Wu \textsuperscript{2}\thanks{ Equal Contribution, $^{\dagger}$ Corresponding Author.}, 
        Hao Li \textsuperscript{3},
        Qinglin Zhang \textsuperscript{2}, \vspace{0.2mm}  \\
        {\bf Yuxiao Hu \textsuperscript{4},}
       {\bf Shuai Zhao \textsuperscript{1 ${\dagger}$},
       {\bf Wenqi Fan \textsuperscript{4 }}}\\
{ 
\textsuperscript{1} Nanyang Technological University, Singapore;
}\vspace{-0.1mm} \\
{ 
\textsuperscript{2} Shanghai Jiao Tong University, China;
}\vspace{-0.1mm} 
{
\textsuperscript{3} Peking University, China;
}\vspace{-0.1mm} \\
{
\textsuperscript{4} Hong Kong Polytechnic University, Hong Kong, China.
}\vspace{-0.1mm}\\
 \texttt{\small shuai.zhao@ntu.edu.sg} \vspace{-0.1mm} \\}

\begin{document}
\maketitle


\begin{abstract}

In AI-facilitated teaching, leveraging various query styles to interpret abstract text descriptions is crucial for ensuring high-quality teaching.
However, current retrieval models primarily focus on natural text-image retrieval, making them insufficiently tailored to educational scenarios due to the ambiguities in the retrieval process.
In this paper, we propose a diverse expression retrieval task tailored to educational scenarios, supporting retrieval based on multiple query styles and expressions.
We introduce the \textbf{S}TEM \textbf{E}ducation \textbf{R}etrieval Dataset~(SER), which contains over 24,000 query pairs of different styles, and the Uni-Retrieval, an efficient and style-diversified retrieval vision-language model based on prompt tuning.
Uni-Retrieval extracts query style features as prototypes and builds a continuously updated Prompt Bank containing prompt tokens for diverse queries.
This bank can updated during test time to represent domain-specific knowledge for different subject retrieval scenarios.
Our framework demonstrates scalability and robustness by dynamically retrieving prompt tokens based on prototype similarity, effectively facilitating learning for unknown queries.
Experimental results indicate that Uni-Retrieval outperforms existing retrieval models in most retrieval tasks.
This advancement provides a scalable and precise solution for diverse educational needs.
\end{abstract}

\section{Introduction}
\label{sec:introduction}
\begin{figure*}[t]
    \centering
    \includegraphics[width=\linewidth]{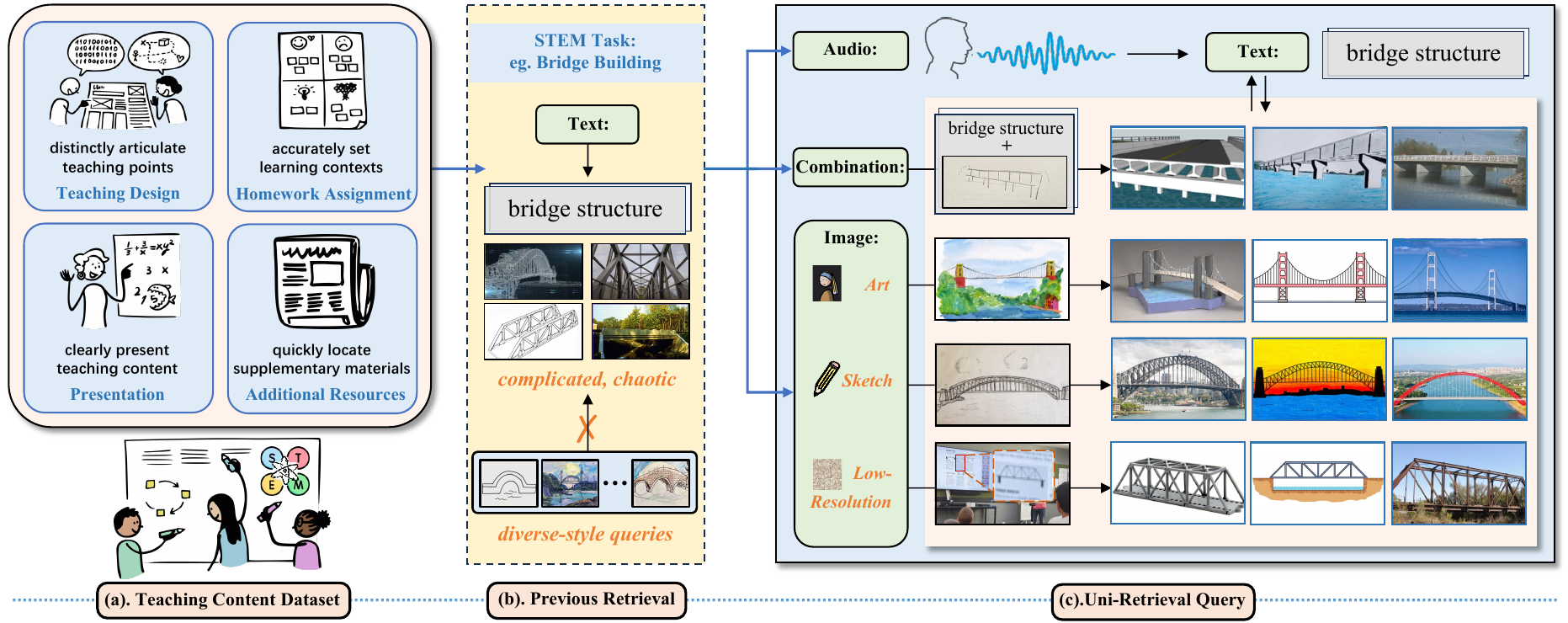}
    \vspace{-5mm}
    \caption{This advancement provides a scalable and precise solution for diverse educational needs. (b). Previous retrieval models focus on text-query retrieval data. (c) Our style-diversified retrieval setting accommodates the various query styles preferred by real educational content.}
    \label{fig:motivation}
    \vspace{-5mm}
\end{figure*}

Artificial Intelligence for Education (AI4EDU) is an emerging interdisciplinary field that leverages AI techniques to transform and enhance instructional design, learning processes, and assessment for education \cite{hwang2020vision}.
With the growing global emphasis on the importance of Science, Technology, Engineering, and Mathematics (STEM) education, retrieving accurate resources from massive interdisciplinary knowledge databases has become a critical challenge. 

Traditional retrieval systems are typically designed for natural text-image contents, which limits their utility in multi-modal STEM education contexts \cite{li2024alleviating,wang2023improving}. Research indicates that these systems often fail to capture the complexity of materials such as images, diagrams, or interactive content, which are vital in STEM disciplines \cite{intro3}. Effective retrieval in STEM education should be able to handle different representations (i.e., different styles of text, image, audio, etc.) to accommodate the diverse learning and teaching needs within STEM subjects. 


Despite advancements in text-image matching techniques \cite{intro12,intro11}, current retrieval systems still encounter challenges when implemented in STEM education \cite{li2025freestyleret}.
These models are primarily optimized for matching text and images, neglecting the variety of query types essential in educational scenarios, including voice, sketches, and low-resolution images \cite{intro6}.  
The constraints of current frameworks within educational contexts frequently result in imprecise, biased, or inefficient retrieval outcomes, such deficiencies can impede teachers' access to suitable instructional resources \cite{intro9}. 


To address above challenges, we propose a multi-style and multi-modal retrieval task tailored for STEM education scenarios, as illustrated in Fig.\ref{fig:motivation}.
The input types for this task include text, audio, and various styles of images, designed to meet the diverse needs of educational contexts.
To adapt to this task, we introduce the STEM Education Multi-Style Retrieval Dataset (SER), curated by 20 graduate and doctoral students from disciplines such as computer science, physics, energy, engineering, and mathematics. The dataset comprises 6,000 natural images with queries in various styles, including sketches, art, low-resolution images, text, and corresponding audio from different STEM fields.
For modeling, we propose a novel plug-and-play feature representation structure called Prompt Bank, which provides a more universal representation of information across different disciplines and semantics by matching the abstract features of data.
Building on Prompt Bank, we develop a lightweight retrieval model named {\bf Uni-Retrieval}.
Uni-Retrieval integrates Prompt Bank with various basic retrieval models, enabling efficient fine-tuning for educational scenarios.
With only a small increase in parameters and inference time, Uni-Retrieval delivers significant performance improvements and fine-tuning capabilities, making it an ideal solution for STEM education retrieval tasks.
Furthermore, it demonstrates the ability to perform effective retrieval across multiple unknown datasets, showcasing its scalability and adaptability in diverse scenarios.
The main contributions of this work can be summarized as follows:
\vspace{-0.5\intextsep}
\begin{itemize}
    \item To bridge the content gap between teacher and student or some abstract expression in STEM education field, we propose the multi-style multi-modal retrieval task for the STEM education. To adapt to this task, we construct the STEM Education Retrieval Dataset (SER), which contains several different subjects in STEM education.
\vspace{-0.5\intextsep}
    \item To efficiently and effectively train a model tailored to our task and dataset, we devise the novel Uni-Retrieval algorithm, which incorporates a sustainably updatable Prompt Bank. Leveraging this bank, the Uni-Retrieval can represent different subjects at a high level and express the features of any other subjects by combining several prompts. 
\vspace{-0.5\intextsep}
    \item Our Uni-Retrieval shows a more strong performance than any other previous method, not only on our SER dataset, but also on traditional retrieval dataset. We believe Uni-Retrieval can bring an infinite potential to STEM education community.
\end{itemize}
\vspace{-0.5\intextsep}

\section{Preliminary}
\subsection{Task Formulation}
We provide a formal problem formulation for query-based retrieval.
Specifically, given an image $I_i$ or a text prompt $P_t$ from the style-specific query set $Q_s$, the retrieval model needs to compute the score between input and target queries and rank the corresponding answers $A$ as high as possible. In the task settings for STEM education retrieval, which share a similar goal, the objective is to rank all answers correctly in response to input queries across various style-specific query sets ${Q}_{s=1}^n$.
If the dataset does not contain the corresponding different style queries, the model should list the same category queries as the suggestions.

\subsection{Dataset Construction}
\begin{figure}[!tbp]
  \centering
   \includegraphics[width=\linewidth]{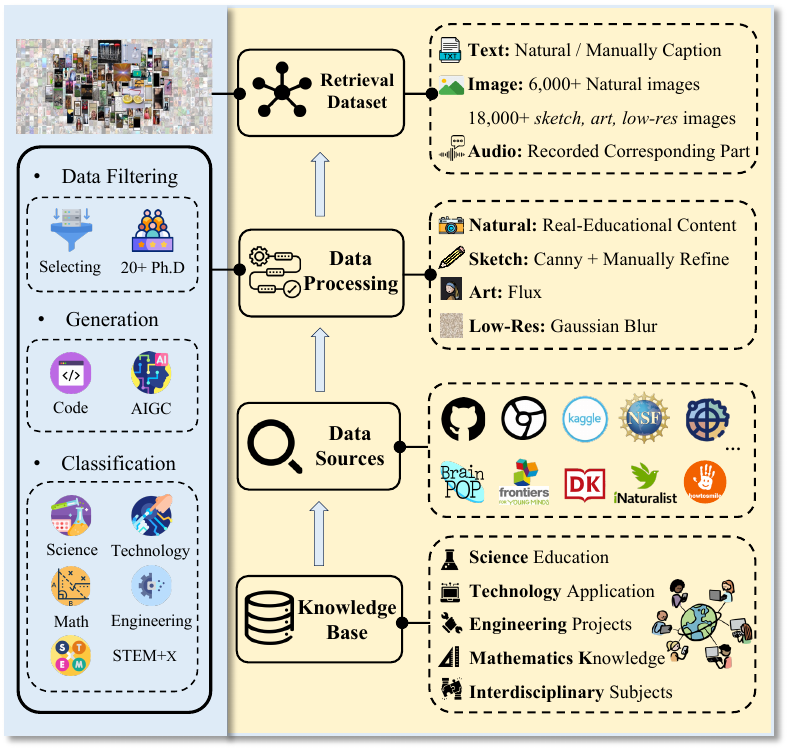}
   \vspace{-7mm}
   \caption{\textbf{Data construction pipeline.} 1. STEM education knowledge base. 2. Data sources: from online resources and dataset researches. 3. Data processing: extracting essential information from collected data, using AIGC algorithms to generate diverse modalities. 4. Retrieval dataset: construct total 24,000 images and multi-modal STEM educational dataset.}
   \label{fig:data_collection}
   \vspace{-5mm}
\end{figure}
SER is a multi-style benchmark dataset we construct to facilitate accurate retrieval for teachers in STEM education. 
It contains a total of 24,000 text captions, audio clips and different style queries to accommodate different educational scenarios.
As illustrated in Fig.\ref{fig:data_sample}, SER contains:
{\bf Text and Natural Image:} the most common query type, allowing teachers to describe problems using natural language or images for retrieval.
{\bf Audio:} a communication medium in education, enabling teachers to articulate complex queries, which can be further enhanced with LLMs or audio encoders.
{\bf Sketch:} hand-drawn sketches, whether created by users or written on blackboards, provide structural cues such as shape, pose, line, and edges to describe the problem.
{\bf Art:} art-style images as queries help bridge the gap between stylistically different images and original images, improving retrieval consistency across styles. 
{\bf Low-Resolution:} queries involving lower-resolution images, such as those captured from a distance, ensure usability in scenarios where high-quality images are unavailable.

The details of the dataset construction pipeline are shown in Fig.\ref{fig:data_collection}, we use the original STEM education image in the source dataset, and extensively collected datasets from the following sources: 1. online resources such as \href{https://www.kaggle.com/datasets}{Kaggle}, \href{https://github.com/carbon-app/carbon}{GitHub}, \href{https://jr.brainpop.com/subject/science/}{BrainPOP}, \href{https://keypoint.keystonesymposia.org/}{Frontiers}, \href{https://learning.dk.com/us}{DKlearning}, \href{https://www.inaturalist.org/}{iNaturalist}, \href{https://www.asiastem.org/steam-free-resources}{AAES}, etc. 2. relevant education dataset research, such as GAN \cite{jin2023gan} and PromptAloud \cite{lee2024prompt}. To ensure high-quality data, more than 20 Ph.D. students from disciplines such as mathematics, physics, chemistry, biology, electronics, computer science, and education conducted a secondary screening of the raw images. They also generate multi-modal combinations (image/text/audio) by leveraging AIGC models.
Based on the natural images, the following steps were undertaken:
{\bf Text Generation:} we manually proofread the natural text descriptions.
{\bf Audio Recording:} the corresponding audio parts are recorded to match the text captions.
{\bf Sketch Images:} Canny algorithms are used to produce sketch images, Pidinet \cite{pdc} is employed to optimize and enhance low-quality sketches, and manual refinement is performed to achieve the final results.
{\bf Art-Style Images:} Flux model \cite{flux} is utilized to create art-style images.
{\bf Low-Resolution Images:} Gaussian Blur algorithms are applied to generate low-resolution images.
According to the National Science Foundation (NSF)’s classification of STEM education, we collect 6,000 original samples spanning six styles, three modalities, and over 22 subjects. This comprehensive dataset, as illustrated in Appendix Fig.\ref{fig:distribution}, ensures diversity and quality across multiple educational domains.

\vspace{-0.5em}
\section{Uni-Retrieval model}
\vspace{-0.5em}

\begin{figure*}[!t]
  \centering
   \includegraphics[width=0.9\linewidth]{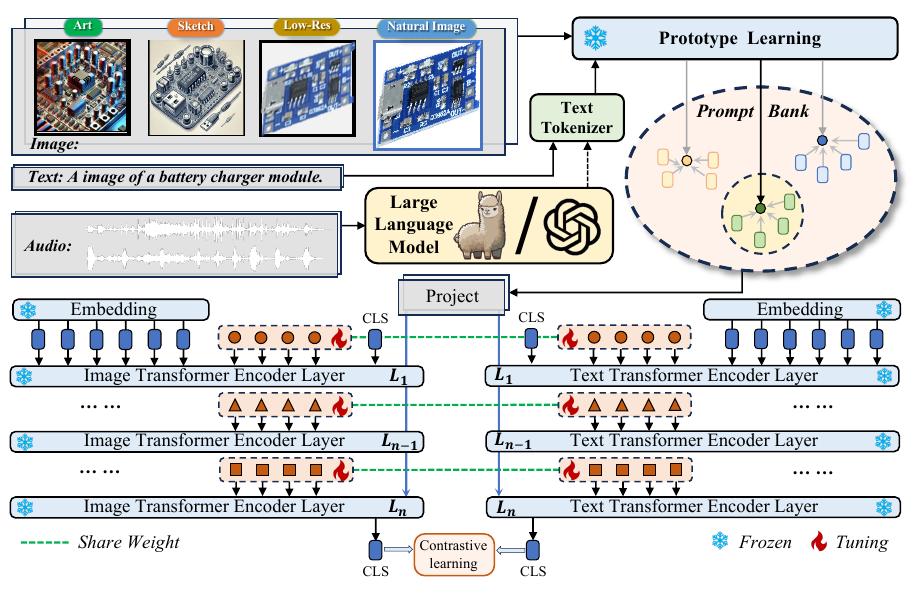}
   \vspace{-3mm}
   \caption{\textbf{The Uni-Retreival model's architechture}.}
   \label{fig:model}
   \vspace{-5mm}
\end{figure*}


Our model consists of three main submodules: a \textbf{prototype learning module} for generating the prototype feature for each content style~(Sec. {\color{red}\ref{subsec:prototype}}), a \textbf{prompt bank} for saving features in the embedding space and selecting the prompt tokens which can represent the input's style~(Sec. {\color{red}\ref{subsec:prompt_bank}}), and a \textbf{feature extractor} based on different vision-language models for retrieving~(Sec.  {\color{red}\ref{subsec:foundation_model}}). Additionally, we further present the training and inference process in Sec. {\color{red}\ref{subsec:training_inference}}.

\subsection{Prototype Learning Module} \label{subsec:prototype}
For Uni-Retrieval, given an input query (freely based on audio, image, or text) \( x \subseteq \mathbb{R}^{L*C} \) and a feature encoder \( f \), map \( x \) into a \( d \)-dimensional shared latent space (Prototype Module) using \( f \), each style has $m$ images. 
For style extracting, researchers usually use the the pretrained models which contains rich semantic information as the feature encoders. For example, if the input queries focus on style images, we leverage the style encoder \cite{styler} to extract visual features. 
If the query emphasizes the need for more textual information from the context, the text encoder and tokenizer, which are pretrained on large text datasets such as the Pile \cite{pile}, can be utilized.
The input query is embedded as follows:
{\setlength{\abovedisplayskip}{5pt}
\setlength{\belowdisplayskip}{5pt}
\begin{eqnarray}
E_{0}^{i} = f(x_{0}^{i}), \quad E_{0}^{i}\subseteq \mathbb R^{d},i=1,2,\dots,m
\end{eqnarray}}

\noindent where $E_{0}^{i}$ denotes the 0-th style's i-th embedding. And then, using the average pooling operation to sum the different each style's embeddings to get the j-th prompt $P_j$:
{\setlength{\abovedisplayskip}{8pt}
\setlength{\belowdisplayskip}{5pt}
\begin{eqnarray}
    P_{j} = \mathrm{AvgPool}(\begin{matrix}\sum^{m}_{i=0}{E_{j}^{i}}\end{matrix})
\end{eqnarray}}

\subsection{Prompt Bank} \label{subsec:prompt_bank}
The Prompt Bank builds a hidden state like TTT \cite{TTT} and mamba \cite{gu2023mamba}, which are designed to store semantic and contextual information at a high level.
Unlike the previous method, which leverages clusters to represent different styles, the Prompt Bank uses hash-like sets to store universal information.
Prompt Bank contains $N$ prompts, and each prompt $k_i$ is associated as a value to a learnable key $P_i$:
{\setlength{\abovedisplayskip}{6pt}
\setlength{\belowdisplayskip}{6pt}
\begin{equation}\setlength{\itemsep}{0.5pt}
Prompt\_Bank \!=\! \{(k_1, P_1),\dots,(k_N, P_N)\}
\end{equation}}

We denote the key set as $K = \{k_i\}^{N}_{i=1}$ and define $\gamma$ to score the match between $k_i$ and $P_i$. Given an input $x$, the $\gamma$ looks up the top-$n$ keys to expand the feature embedding of $x$. The aim to use the hash-liked structure is promoting the matching speed between the input and the Prompt Bank's tokens. For Uni-Retrieval model, we calculate the cosine similarity between the matching prompt $P_{j_i}$ and the key $K_i$:
{\setlength{\abovedisplayskip}{7pt}
\setlength{\belowdisplayskip}{7pt}
\begin{eqnarray}
K_x={\underset{\{j_i\}^{n}_{i=1}\subseteq [1,N]}{\arg\min}}\quad \begin{matrix}\sum^{n}_{i=1}\gamma(P_{j_i}, K_i)\end{matrix}
\end{eqnarray}}

The Prompt Bank is free to combine different prompt tokens and expand feature embedding space, allowing different tokens associated with specific styles to jointly represent an input query. Due to the generalization properties on out-of-distribution of our Prompt Bank, even unseen styles also can combine similar styles' tokens to enhance retrieval performance by expand the semantic/context information provided by Prompt Bank. The expanding method is suitable for both different styles of images and different expression of text. 
Usually the special token is put at the beginning of the sequence tokens:
{\setlength{\abovedisplayskip}{6pt}
\setlength{\belowdisplayskip}{6pt}
\begin{eqnarray}
x_p = [CLS;P_{j_1};P_{j_2};\dots;P_{j_n};x_e]
\end{eqnarray}}

\noindent where $x_p$ denotes the image's input feature after expanding prompt tokens; $x_e$ represents  the original sequence tokens patched from the input; $CLS$ is the special token used for performing downstream tasks with different heads.


\subsection{Feature Extractor}
\label{subsec:foundation_model}
In Uni-Retrieval, we apply the ViT structure as the visual feature extractor, leverage a tokenizer to embed the input text query $x \subseteq \mathbb{R}^{L*C}$, and then utilize the transformer \cite{vaswani2017attention} as the text encoder to extract features. 
The vision encoder and the text encoder are initialized with OpenCLIP, and gpt-neo \cite{gpt-neo} trained on the Pile dataset as the tokenizer, respectively.
What's more, we uses GPT-4o \cite{hurst2024gpt} as the optional choice of the external large language model to convert the audio clips to the text sequence.

The whole sequence tokens are feed into the feature extractor for training and inference layer by layer. Obtained from the Prompt Bank, visual prompt tokens represent various style information specific to different STEM subjects, while text prompt tokens convey distinct context information about different STEM subjects or different expression about the same subjects. 
The parameters are sharing between visual prompts and text prompts each layer to align vision and text modality. 
In the model training phase, the vision encoder, the tokeniser and the text encoder are fully frozen to retain the original semantic space.

\subsection{Training and Inference}
\label{subsec:training_inference}

For the training procedure, in every training step, the style/context features are extracted from the corresponding encoder $f$ to get the prompt $P_j$. Then, matching $P_j$ and the key $K_j$ to get the matching prompts $P_j$. Besides, the tokenizer and the patch layer map the inputs into sequence $x_t$: 
{\setlength{\abovedisplayskip}{6pt}
\setlength{\belowdisplayskip}{6pt}
\begin{eqnarray}
x_t = \mathrm{Tokenizer/Patch}(x)
\end{eqnarray}}

\noindent where $x_t$ denotes the temp state of features. 
After selecting $n$ prompts following the aforementioned query strategy, the expanded feature embedding $x_p$ is fed into the foundation model $\delta$ and getting the final result $x_f$. We use the $CLS$ token to represent the whole sequence $x_p$ following the settings of LLaMA \cite{touvron2023llama}:
{\setlength{\abovedisplayskip}{6pt}
\setlength{\belowdisplayskip}{6pt}
\begin{eqnarray}
x_f = \delta(x_p)[:, 0, :]
\end{eqnarray}}

\noindent The triplet loss function $\mathcal{L}$ utilizes the features $x_f$, $x_r$, and $x_h$ of an image or text, a retrieval target query, and a negative sample from a different category. 
Minimizing $\mathcal{L}$ brings the correct sample pairs closer together while distancing the negative sample pairs.
With $\mu$ as margin and distance function $d(a,b)=(1-a*b)/(||a||-||b||)$, $\mathcal L$ is given as: 
{\setlength{\abovedisplayskip}{6pt}
\setlength{\belowdisplayskip}{6pt}
\begin{multline}
\mathcal L = max\{0, \\
\mu + d(\delta(x_f), \delta(x_r))-d(\delta(x_f), \delta(x_h))\}
\end{multline}}

\noindent where $x_r, x_h$ denotes the embedding of the retrieval object and the negative sample respectively. Moreover, the key in Prompt Bank will be updated with a scale parameter $\lambda$ to weight the second loss function term:
{\setlength{\abovedisplayskip}{6pt}
\setlength{\belowdisplayskip}{6pt}
\begin{eqnarray}
\min_{k,p,L} \mathcal L(x_f, x_r, x_h) + \lambda \begin{matrix}\sum_{K_x}\gamma(q(x), k_{si})\end{matrix}
\end{eqnarray}}

During the training procedure, Uni-Retrieval will jointly train both the Prompt Bank's keys $K$ and the prompt tokens $P$, which endow the Prompt Bank with the capability for continuous updates. The updatable parameters of the Uni-Retrieval model are limited compared to full-parameter fine-tuning, effectively saving computational resources and enhancing training speed. 

For the inference procedure, we first extract the prototype feature for the unknown-style query input $x$. We use the prototype feature as the query to retrieve the fittest prompt tokens from the prompt bank. If the style is unknown type, the Prompt Bank will use several different clusters to represent it jointly. Then we prepend the prompt tokens $x_p$ to the feature extractor $\delta$ for retrieval. The query embeddings are extracted in advanced and stored in the databases to accelerate the retrieval process.

\section{Experiments}

\subsection{Experiments Settings}

\newcommand{\pub}[1]{\color{red}{\tiny{#1}}}
\newcommand{\Frst}[1]{{\textbf{#1}}}
\newcommand{\Scnd}[1]{{\underline{#1}}}
\begin{table*}[htp]
\centering
\footnotesize
\renewcommand{\arraystretch}{1.05}  
\setlength{\tabcolsep}{2.75mm}        
{
{
\begin{tabular}{c|c|cc|cc|cc|cc}
    \toprule[1.5pt]
    \multirow{2}{*}{\textbf{\#}} & \multirow{2}{*}{\textbf{Method}} & \multicolumn{2}{c|}{\textbf{Text} \textbf{$\rightarrow$} \textbf{Image}} & \multicolumn{2}{c|}{\textbf{Sketch} \textbf{$\rightarrow$} \textbf{Image}} & \multicolumn{2}{c|}{\textbf{Art} \textbf{$\rightarrow$} \textbf{Image}} & \multicolumn{2}{c}{\textbf{Low-Res} \textbf{$\rightarrow$} \textbf{Image}} \\ 
    
    \cmidrule(rl){3-4}\cmidrule(rl){5-6}\cmidrule(rl){7-8}\cmidrule(rl){9-10}
    & & {R@1$\uparrow$} & {R@5$\uparrow$} & {R@1$\uparrow$} & {R@5$\uparrow$} & {R@1$\uparrow$} & {R@5$\uparrow$} & {R@1$\uparrow$} & {R@5$\uparrow$} \\

    \noalign{\hrule height 1.5pt}
    \rowcolor{gray!20}\multicolumn{10}{c}{\it{\textbf{Pretrained Cross-Modality Models}}} \\
    \hline
    1& CLIP          & 54.6 & 78.4 & 47.3 & 68.9 & 46.8 & 71.3 & 53.7 & 72.9\\
    2& BLIP          & 55.8 & 79.2 & 48.2 & 69.2 & 47.5 & 74.4 & 51.5 & 74.2\\
    3& CLIP-Finetune                & 71.4 & 91.4 & 71.0 & 87.0 & 52.2 & 81.6 & 71.2 & 88.1\\
    4& BLIP-Finetune                & 70.2 & 92.0 & 71.6 & 89.2 & 54.3 & 82.3 & 69.7 & 86.8\\
    \hline
    \rowcolor{gray!20}\multicolumn{10}{c}{\it{\textbf{Large Multi-Modality Models}}} \\
    \hline
    5& LanguageBind     & 60.2 & 86.9 & 52.8 & 78.4 & 49.0 & 78.4 & 59.1 & 80.2\\
    6& Unified-IO2   & 67.5 & 89.2 & 59.6 & 84.1 & 55.9 & 82.9 & 64.3 & 84.0\\
    \hline
    \rowcolor{gray!20}\multicolumn{10}{c}{\it{\textbf{Style Retrieval Models}}} \\
    \hline
    7& SceneTrilogy  & 69.7 & 84.5 & 75.6 & \textbf{96.5} & 71.5 & 92.9 & 68.6 & 85.5\\
    8& FashionNTM    & 50.4 & 81.3 & 68.9 & 88.6 & 67.1 & 88.9 & 45.6 & 77.5\\
    \hline
    \rowcolor{gray!20}\multicolumn{10}{c}{\it{\textbf{Cross-Modality Prompt Learning Models}}} \\
    \hline
    9& VPT          & 69.9 & 84.1 & 53.3 & 72.3 & 62.7 & 83.2 & 67.4 & 79.1\\
    10& CoCoOP       & 72.2 & 86.7 & 53.8 & 74.8 & 66.4 & 87.4 & 70.8 & 81.6\\
    11& MaPLe       & 73.8 & 87.8 & 62.7 & 78.9 & 67.8 & 89.4 & 71.9 & 86.3\\
    12& FreestyleRet & 80.1 & 92.5 & 75.3 & 91.5 & 73.0 & \textbf{98.3} & 78.0 & 90.7\\
    \hline
    \rowcolor{gray!20}\multicolumn{10}{c}{\it{\textbf{Database-Driven Retrieval Models}}} \\
    \hline
    13& GASKN       & 55.7 & 80.8 & 47.6 & 68.7 & 48.5 & 75.9 & 53.6 & 70.5\\
    14& SKG         & 57.8 & 82.1 & 45.4 & 65.3 & 49.2 & 76.1 & 56.8 & 75.4\\
    \noalign{\hrule height 1pt}
    \rowcolor{aliceblue!60} 15& \textbf{Uni-Retrieval}  & \textbf{83.2} & \textbf{98.7} & \textbf{84.5} & 95.6 & \textbf{76.9} & 97.5 & \textbf{87.4} & \textbf{98.1}\\ 
 \bottomrule[1.5pt]
\end{tabular}
}
}
\vspace{-2mm}
\caption{Retrieval performance for STEM Education Retrieval task.}
\label{tab:main_results}
\vspace{-3mm}
\end{table*}

We use our SER as the main experiment analysis and another three retrieval datasets to comprehensively evaluate the Uni-Retrieval's performance. For evalution metric, We evaluate the R@1 and R@5 metrics and the inference speed~(ms) on all retrieval datasets. For R@1 and R@5, ``$\uparrow$'' denotes that higher is better. For ms, ``$\downarrow$'' denotes that quicker is better.
Implement Details are detailed in Appendix ~\ref{supsubsec:Experiment Settings}.

\begin{table}[htp]
    \centering
    \resizebox{1.0\columnwidth}!{
    \begin{tabular}{c|cccc}
        \toprule[1.5pt]
        \textbf{Method} & \textbf{Params(M)} & \textbf{Q2I(ms)$\downarrow$} & \textbf{Q2T(ms)$\downarrow$} & \textbf{T$\rightarrow$I(Acc)}$\uparrow$ \\
        \hline
        CLIP & 427M & 68ms & 63ms & 54.6\\
        BLIP & 891M & 62ms & 58ms & 55.8\\
        VPT & 428M & 73ms & 69ms & 69.9\\
        LanguageBind & 1200M & 372ms & 367ms & 60.2\\
        GASKN & 33M & 12ms & 10ms & 55.7\\
        \hline
        \rowcolor{aliceblue!60} \textbf{Uni-Retrieval} & 453M{$_{\textcolor{red}{(+26)}}$} & 77ms{$_{\textcolor{red}{(+9)}}$} & 70ms{$_{\textcolor{red}{(+7)}}$} & 83.2{$_{\textcolor{red}{(+28.6)}}$}\\ 
        \bottomrule[1.5pt]
        \hline
    \end{tabular}}
    \vspace{-2mm}
    \caption{The models inference speed comparison.}
    \label{tab:speed}
    \vspace{-5mm}
\end{table}

\begin{table}[!h]
    \centering
    \resizebox{1.0\columnwidth}!{
    \begin{tabular}{c|cccc}
        \toprule[1.5pt]
        Method & T$\rightarrow$I & T+S$\rightarrow$I & I$\rightarrow$T & I+S$\rightarrow$T \\
        \hline
        CLIP-Finetune & 54.6 & 55.3{$_{\textcolor{red}{(+0.7)}}$} & 47.4 & 46.6{$_{\textcolor{green}{(-0.8)}}$}\\
        VPT & 69.9 & 72.0{$_{\textcolor{red}{(+2.1)}}$} & 73.9 & 74.1{$_{\textcolor{red}{(+0.2)}}$}\\
        \hline
        \rowcolor{aliceblue!60} \textbf{Uni-Retrieval} & 83.2 & 87.4{$_{\textcolor{red}{(+4.2)}}$} & 81.7 & 83.3{$_{\textcolor{red}{(+1.6)}}$}\\ 
        \bottomrule[1.5pt]
    \end{tabular}}
    \vspace{-2mm}
    \caption{Retrieval performance with multi-style queries simultaneously on SER dataset.}
    \label{tab:multi-query}
    \vspace{-5mm}
\end{table}

\begin{table*}[!tbp]
\centering
\footnotesize
\renewcommand{\arraystretch}{1.05}  
\setlength{\tabcolsep}{2.25mm}        
{
\begin{tabular}{c|c|cccc|ccc|cc|c}
    \toprule[1.5pt]
    
    \textbf{\#} & \textbf{Method} & \textbf{I$\rightarrow$T} & \textbf{S$\rightarrow$T} & \textbf{A$\rightarrow$T} &  \textbf{L$\rightarrow$T} & \textbf{T$\rightarrow$S} & \textbf{T$\rightarrow$A} & \textbf{T$\rightarrow$L} & \textbf{S$\rightarrow$A} & \textbf{S$\rightarrow$L} & \textbf{A$\rightarrow$L} \\
    

    \noalign{\hrule height 1.5pt}
    \rowcolor{gray!20}\multicolumn{12}{c}{\it{\textbf{Metric: R@1$\uparrow$ on SER Dataset}}} \\
    \hline
    1& CLIP          & 47.4 & 38.4 & 37.9 & 38.6 & 38.8 & 37.4 & 35.7 & 36.9 & 34.8 & 31.5 \\
    2& BLIP          & 48.9 & 39.2 & 38.4 & 39.5 & 39.7 & 37.1 & 36.5 & 35.0 & 34.9 & 32.6 \\
    3& CLIP-Finetune                & 75.7 & 72.4 & 71.3 & 69.8 & 70.2 & 68.5 & 67.7 & 65.4 & 66.8 & 66.3\\
    4& BLIP-Finetune                & 77.4 & 73.0 & 72.6 & 70.5 & 71.3 & 69.4 & 68.1 & 66.2 & 67.2 & 67.0\\
    \hline
    5& LanguageBind     & 55.4 & 54.9 & 53.1 & 53.4 & 49.7 & 48.7 & 49.1 & 46.2 & 46.8 & 45.9\\
    6& Unified-IO2   & 57.3 & 57.2 & 56.3 & 54.5 & 51.1 & 49.9 & 48.6 & 48.0 & 47.2 & 46.8\\
    \hline
    7& SceneTrilogy  & 72.4 & \textbf{76.5} & 70.6 & 71.5 & 69.3 & 69.9 & 68.7 & 65.2 & 66.2 & 64.4 \\
    8& FashionNTM   & 70.6 & 73.3 & 68.9 & 69.6 & 67.1 & 68.0 & 66.5 & 67.5 & 64.8 & 62.4 \\
    \hline
    9& VPT          & 73.9 & 71.8 & 70.4 & 68.7 & 69.0 & 68.2 & 67.4 & 66.6 & 64.5 & 63.8\\
    10& CoCoOP       & 76.5 & 74.7 & 73.4 & 74.0 & 71.4 & 72.3 & 70.8 & 68.9 & 67.2 & 67.3\\
    11& MaPLe      & 78.3 & 75.8 & 75.7 & 74.9 & 72.4 & 69.6 & 69.2 & 68.3 & 67.4 & 65.6 \\
    12& FreestyleRet & 80.8 & 73.5 & \textbf{75.5} & 71.4 & 73.0 & 68.3 & 68.0 & 69.4 & 70.6 & 68.9 \\
    \hline
    13& GASKN        & 53.8 & 52.9 & 52.6 & 50.7 & 49.4 & 47.9 & 46.0 & 47.1 & 47.3 & 45.9\\
    14& SKG          & 54.3 & 51.7 & 50.4 & 51.3 & 48.5 & 46.1 & 45.4 & 46.9 & 47.0 & 45.9\\
    \noalign{\hrule height 1pt}
    \rowcolor{aliceblue!60} 15& \textbf{Uni-Retrieval}  & \textbf{81.7} & 76.3 & \textbf{74.9} & \textbf{77.6} & \textbf{73.5} & \textbf{74.2} & \textbf{78.0} & \textbf{71.4} & \textbf{72.3} & \textbf{70.8}\\ 
 \bottomrule[1.5pt]
\end{tabular}
}
\vspace{-2mm}
\caption{Retrieval performance for STEM Education Retrieval task.}
\label{tab:other_results}
\vspace{-5mm}
\end{table*}

\subsection{Comparison Experiment}

On the SER dataset, Uni-Retrieval demonstrates superior performance across multiple scenarios with different query styles compared to other baselines, including multi-modality models, cross-modality pre-trained models, and prompt learning models. As shown in Tab.\ref{tab:main_results} and Tab.\ref{tab:other_results}, the $T+S\!\rightarrow\!I$ means inputting the text and the style images as the multi-queries and outputting the corresponding images as the target queries. The experiment results yield three key observations:

\noindent \textbf{The Uni-Retrieval achieves the best retrieval performance on the multi-style STEM Education Retrieval task:} This highlights the effectiveness of Uni-Retrieval in handling complex multi-modal queries. Due to the Prompt Bank's structure, Uni-Retrieval is a plug-and-play framework that can be highly flexible applied to various multi-modal models and enhance their retrieval capabilities. Line 15 in Tab.\ref{tab:main_results} provides a substantial performance boost compared to both CLIP and CLIP-Finetune, further validating the effectiveness of our framework.

\noindent \textbf{The Prototype module and Prompt Bank significantly outperform full-parameter fine-tuning:} As shown in lines 3-4 and line 15 of Tab.\ref{tab:main_results}, Uni-Retrieval surpasses its fine-tuned CLIP counterpart by a large margin. Leveraging the prior knowledge bias introduced by the Prototype module and the efficient memory space of the Prompt Bank, Uni-Retrieval achieves superior results while tuning less than 5\% of the model’s total parameters. This demonstrates the effectiveness of Uni-Retrieval’s design in achieving high performance with minimal parameter adjustments.

\noindent \textbf{Uni-Retrieval can simultaneously perform and mutually enhance traditional text-image retrieval performance:} As shown in Tab.\ref{tab:multi-query}, when handling text-image retrieval tasks, Uni-Retrieval allows multi-query inputs as additional references, significantly improving retrieval capability. This multi-query input design is not exclusive to Uni-Retrieval and can also benefit other retrieval models, offering a generalizable approach to enhancing retrieval tasks.

In addition to accuracy, inference speed is a crucial metric for evaluating retrieval models. As shown in Tab.\ref{tab:speed}, Uni-Retrieval adds just 9ms per search iteration. 
Compared to GASKN, Uni-Retrieval demonstrates significantly stronger retrieval performance than traditional database-driven methods. 
Additionally, when compared to other cross-modality methods, Uni-Retrieval excels in both tuning efficiency and retrieval accuracy, further validating its effectiveness and scalability.

\begin{table}[h]
    \centering
    
    \resizebox{1.0\columnwidth}!{
    \begin{tabular}{c|c|c|cccc}
        \toprule[1.5pt]
        \textbf{\#} & \textbf{Type} & \textbf{Token-Num} & \textbf{T\textbf{$\rightarrow$}I} & \textbf{S\textbf{$\rightarrow$}I} & \textbf{A\textbf{$\rightarrow$}I} & \textbf{L\textbf{$\rightarrow$}I} \\ 
        \hline
        1 & Deep & 1 & 72.0 & 78.3 & 73.2 & 80.6\\
        2 & Deep & 2 & 77.1 & 81.2 & 75.5 & 85.8\\
        3 & Deep & 8 & \textbf{83.24} & 82.7 & 76.5 & 87.0\\
        4 & Shallow & 4 & 68.2 & 75.6 & 70.4 & 77.3\\
        \hline
        \rowcolor{aliceblue!60} 
        5 & Deep & 4 & 83.2 & \textbf{84.5} & \textbf{76.9} & \textbf{87.4}\\
        \bottomrule[1.5pt]
    \end{tabular}
    }
    \vspace{-2mm}
    \caption{Ablation study for prompt settings.}
    \label{ablation:prompt}
    \vspace{-5mm}
\end{table}

\subsection{Ablation Study}

To quantitatively evaluate the role of prompts in the model, we conducted ablation studies on the insertion type and token number of prompt tokens within Uni-Retrieval for four style metrics. These studies aimed to assess their impact on the style-diversified STEM education retrieval task, providing insights into how prompts influence performance and model adaptability. The shallow type involves inserting prompt tokens only at the first layer, while the deep type inserts prompt tokens across all layers. The token number refers to the number of repeated prompts. 

As shown in lines 4-5 in Tab.\ref{ablation:prompt}, the results indicate that the deep type consistently outperforms the shallow type. Additionally, lines 1-3 and line 5 in Tab.\ref{ablation:prompt} demonstrate the change on number of prompt tokens. We observed that repeating the prompt tokens more than four times does not significantly improve model performance. Instead, it rapidly increases the number of tuning parameters, which slows down both the tuning process and inference speed. This indicates that four repetitions provide a balanced trade-off between performance and computational efficiency. Therefore, we ultimately selected four prompt tokens to be inserted at each layer as the standard configuration for Uni-Retrieval. This choice effectively balances model performance, tuning efficiency, and inference speed, which can serve as a valuable reference for other tuning models.

\begin{table}[h]
\centering
\footnotesize
\renewcommand{\arraystretch}{1.05}  
\setlength{\tabcolsep}{0.5mm}        
{
{
\begin{tabular}{c|cc|cc|cc}
    \toprule[1.5pt]
    \multirow{2}{*}{\textbf{Method}} & \multicolumn{2}{c|}{\textbf{Text} \textbf{$\rightarrow$} \textbf{Im.}} & \multicolumn{2}{c|}{\textbf{Sketch} \textbf{$\rightarrow$} \textbf{Im.}} & \multicolumn{2}{c}{\textbf{Art} \textbf{$\rightarrow$} \textbf{Im.}}  \\ 
    
    \cmidrule(rl){2-3}\cmidrule(rl){4-5}\cmidrule(rl){6-7}
    & {R@1$\uparrow$} & {R@5$\uparrow$} & {R@1$\uparrow$} & {R@5$\uparrow$} & {R@1$\uparrow$} & {R@5$\uparrow$} \\

    \noalign{\hrule height 1.5pt}
    \rowcolor{gray!20}\multicolumn{7}{c}{\it{\textbf{Diverse-Style Retrieval Dataset}}} \\
    \hline
    LanguageBind     & 71.0 & 95.5 & 50.8 & 79.4 & 58.2 & 86.3 \\
    CoCoOP~       & 71.4 & 94.6 & 77.5 & 97.2 & 69.3 & 97.1 \\
    MaPLe         & 73.1 & 95.9 & 80.3 & 97.9 & 70.6 & 97.2 \\
    FreestyleRet  & 71.4 & 97.2 & 81.6 & 98.0 & 72.3 & \textbf{98.1} \\
    \hline
    \rowcolor{aliceblue!60}
    Uniretrieval          & \textbf{82.3} & \textbf{97.4} & \textbf{82.7} & \textbf{98.9} &                                             \textbf{75.1} & 98.0 \\
    \hline
    \rowcolor{gray!20}\multicolumn{7}{c}{\it{\textbf{DomainNet Dataset}}} \\
    \hline
    VPT          & 59.7 & 86.1 & 53.5 & 77.3 & 54.6 & 81.8 \\
    BLIP-Finetune               & 65.3 & 94.2 & 71.4 & 89.7 & 54.3 & 82.3 \\
    FreestyleRet & 70.2 & 95.2 & 75.2 & 93.2 & 73.1 & 92.6 \\
    \hline
    \rowcolor{aliceblue!60}
    Uniretrieval         & \textbf{70.7} & \textbf{96.0} & \textbf{77.6} & \textbf{94.1} &                                             \textbf{73.4} & \textbf{92.9} \\
    \hline
    \rowcolor{gray!20}\multicolumn{7}{c}{\it{\textbf{SketchCOCO Dataset}}} \\
    \hline
    MaPLe        & 26.4 & 53.9 & 18.0 & 48.3 & - & - \\
    SceneTrilogy & 30.6 & 65.8 & 22.5 & 51.6 & - & - \\
    FreestyleRet & 31.5 & 67.3 & 29.6 & 56.1 & - & - \\
    \hline
    \rowcolor{aliceblue!60}
    Uniretrieval         & \textbf{34.7} & \textbf{71.6} & \textbf{30.2} & \textbf{60.4} & - & - \\
 \bottomrule[1.5pt]
\end{tabular}
}
}
\vspace{-2mm}
\caption{The zero-shot retrieval performance comparison on retrieval datasets.}
\vspace{-5mm}
\label{tab:multi_results}
\end{table}

\begin{figure*}[h]
  \centering
   \includegraphics[width=0.95\linewidth]{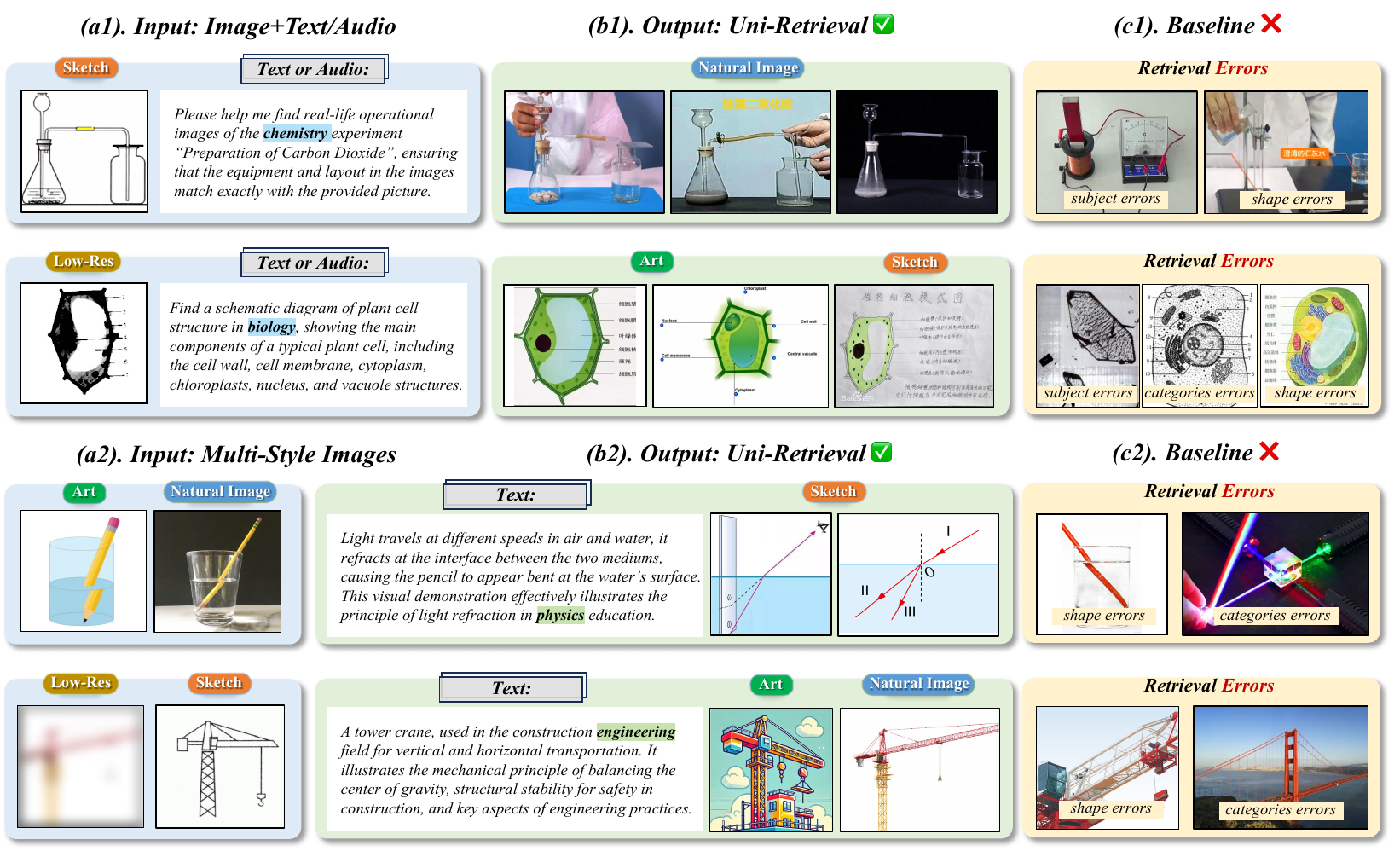}
   \vspace{-2mm}
   \caption{The case study for our Uni-Retrieval and the FreestyleRet baseline.}
   \vspace{-5mm}
   \label{fig:visualization}
\end{figure*}

We also evaluated Uni-Retrieval’s zero-shot retrieval performance on several other multi-style datasets. As shown in Tab.\ref{tab:multi_results}, we compared Uni-Retrieval against various baseline models across three datasets: the DSR, DomainNet, and SketchCOCO dataset, each representing distinct domains of style-based queries. 
As shown in Tab.\ref{tab:multi_results}, Uni-Retrieval demonstrates exceptional zero-shot performance across these diverse datasets, highlighting its capability to perform effective information retrieval in various previously unknown databases.
This performance underscores Uni-Retrieval’s scalability and robustness, significantly enhancing the adaptability and effectiveness of existing retrieval models in handling diverse and unstructured data domains.

\subsection{Visualization Result}

In Fig.\ref{fig:visualization}, we visualize the style-diversified query inputs and their corresponding retrieval answers from our Uni-Retrieval and the FreestyleRet baseline model. We summarize three common retrieval errors in the case analysis, where subject errors, semantic errors, and color errors represent the false retrieval result with false subjects, semantic information, and colors. We propose the subject error cases in Fig.\ref{fig:visualization}(a1)-(c1). The subject information is contained widely in different style queries. Thus, pose error cases occur in sketch, art, and low-resolution queries. 
Subject information is conveyed through the primary objects in images and their corresponding textual descriptions. Subject errors occur when there is incorrect recognition or classification of these objects, leading to mismatched associations between the image and text.

Semantic errors, on the other hand, arise from inaccuracies in describing object details. These errors frequently occur when irrelevant text is associated with specific parts of an object, particularly in the context of art descriptions. Such mismatches result in the model generating incorrect attention maps, thereby failing to accurately connect the visual and textual elements. Thus, in Fig.\ref{fig:visualization}(a2)-(c2), most of the semantic errors occur in the art-style retrieval task. 

For the low-resolution query retrieval task, color is vital retrieval information. We show the color errors from the low-resolution retrieval task.  Compared to the FreestyleRet baseline model, our Uni-Retrieval achieves fine-grained retrieval based on subject, semantic, and color information from style-diversified query inputs. It demonstrates a superior understanding of semantic information and fine-grained alignment between modalities, particularly in the precise description and representation of key object parts. This highlights the significant advantages and capabilities of our Uni-Retrieval framework.
\section{Conclusion}
To address the challenge of fine-grained and efficient retrieval in STEM teaching scenarios, we proposed a multi-style and multi-modal STEM education retrieval task and curated a multi-style dataset of over 24,000 samples for model fine-tuning.
To balance training efficiency and retrieval performance, we developed a lightweight and plug-and-play feature expression module, Prompt Bank, and built a database-driven accurate retrieval model, Uni-Retrieval, based on Prompt Bank.
Compared to current state-of-the-art retrieval models, Uni-Retrieval significantly improves retrieval performance with only a 26M (less than 5\%) increase in parameter size and less than 10ms additional retrieval time. Furthermore, the training and deployment costs of Uni-Retrieval are substantially lower than those of existing large retrieval models, making it a more economical and practical solution for educational scenarios.
We hope Uni-Retrieval can inspire new possibilities for the community, offering an effective and accessible approach to retrieval in STEM education and beyond.

\section*{Limitation}
However, our work still has some limitations that require further research. Firstly, STEM education differs significantly from higher education, K-12 education, humanities education, and other scenarios in terms of data and usage requirements. A key challenge for future research is how to maintain efficient retrieval performance while adapting to a wider range of educational scenarios.
Additionally, we plan to exploring how to efficiently acquire various professional educational knowledge based on VLMs. These improvements aim to make Uni-Retrieval more versatile and impactful across diverse educational domains.

\bibliography{custom}

\clearpage

\appendix

\section{Related Works}
\label{sec:related_work}

\subsection{Dataset Adaptation in Education}
Within the realm of education, image retrieval in education has distinct characteristics, as images often reflect the teaching intentions of educators. This facilitates the rapid and accurate alignment of visual content with teaching materials, thereby reducing educators' preparation workload and enhancing the precision of learning data. While existing researches has focused on classifying educational data \cite{choi2020ednet}, they often encounter constraints. Due to the complexity of incorporating an expansive range of teaching scenarios in STEM education and the scarcity of data, numerous studies often narrow the scope to a limited set of subject applications \cite{hendrycks2021measuring,pal2022medmcqa} or to a limited set of teaching strategy retrievals \cite{kwon2024biped,welbl2017crowdsourcing}.

\begin{figure*}[!t]
    \centering
    \includegraphics[width=\linewidth]{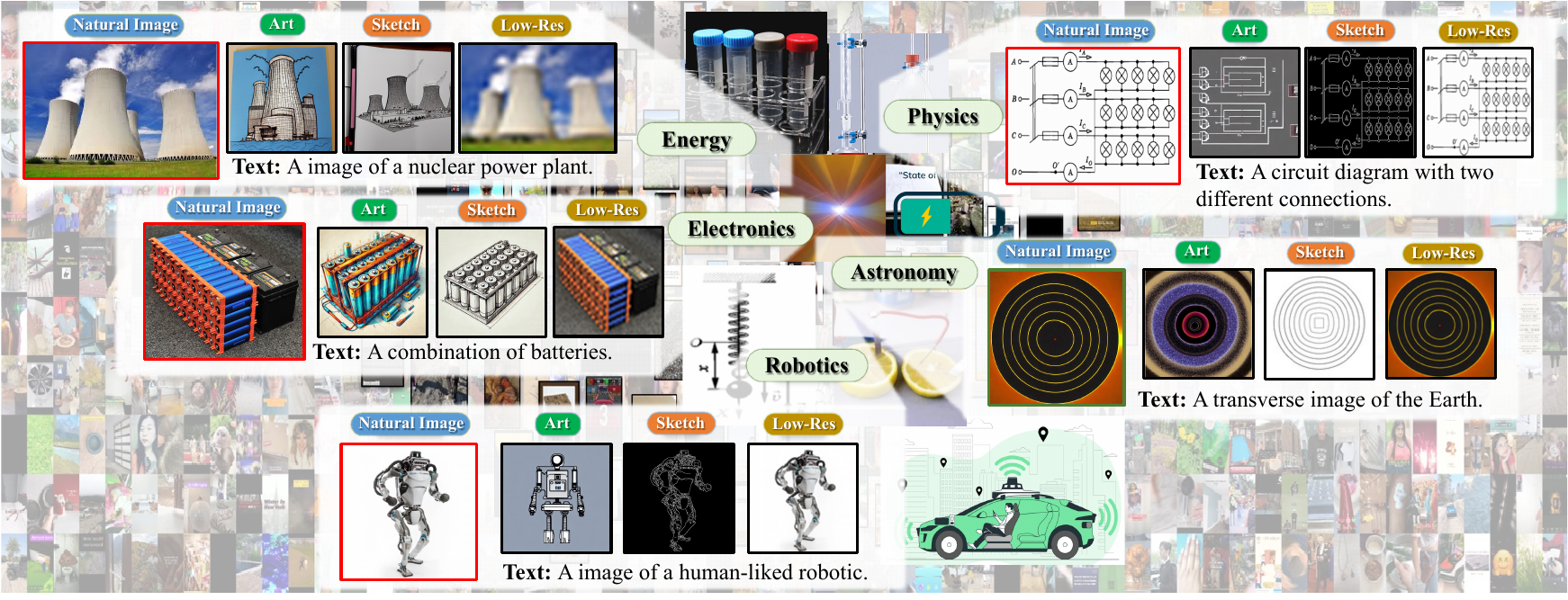}
    \vspace{-7mm}
    \caption{\textbf{The SER Dataset} contains 24,000+ text captions and their corresponding queries with various styles, including Natural, Sketch, Art, Low-Resolution~(Low-Res) images and audio clips from different STEM subjects.}
    \vspace{-5mm}
    \label{fig:data_sample}
\end{figure*}

There is a considerable variation across existing STEM education datasets regarding its specific composition. Many datasets are cluttered with irrelevant or invalid data, lack comprehensive coverage of specialised content, and suffer from quality assurance issues \cite{patrinos2018global}. 
Although STEM education datasets are assembled from interactions between learners and large language models \cite{hou2024eval, wang2024large}, they are generally not well-suited for use by educators and learners across multiple domains. Furthermore, creating a precise and professional data retrieval repository for the educational domain requires efficient retrieval algorithms as support \cite{alzoubi2024enhancing}. To ensure efficient retrieval and usability in STEM education scenarios, we construct the SER dataset, which includes multiple query styles to enhance retrieval diversity.

\subsection{Multi-task Learning}

In STEM education, the multi-style retrieval model needs to leverage multi-task learning to align features and learning across different modal samples. Multi-task learning refers to the simultaneous training and optimization of multiple related tasks within a single model \cite{9392366,xiao2025exploring,jia2025seeing,jia2025robust}. By sharing parameters and representations across functions, it improves overall performance. Compared to other transfer learning methods, including domain adaptation \cite{farahani2021brief} and domain generalization \cite{zhou2022domain}, multi-task learning annotates data and achieves CLIP-level model fine-tuning and convergence, the data of each task in multi-task learning is well-labeled. 

Overall, multi-task learning introduces a new tool for STEM education practitioners that may help meet requirements, especially if speed and efficiency are preferred over performance. While many recent multi-task learning employ two clusters of contemporary techniques, hard parameter sharing and soft parameter sharing \cite{ruder2017overview}. In hard parameter sharing, most or all of the parameters in the network are shared among all tasks \cite{kokkinos2017ubernet}. In soft parameter sharing,the models are tied together either by information sharing or by requiring parameters to be similar \cite{yang2016trace}. Consequently, our Uni-Retrieval adopts a blended multi-task learning paradigm, adopt both hard and soft  parameter in different styles of tasks. Building upon successul multi-task learning method for CLIP, such as CoCoOP \cite{zhou2022conditional}, MaPLe \cite{khattak2023maple}, and FreestyleRet \cite{li2025freestyleret}, our study leverages these techniques to strengthen domain adaptation and multi-task learning. 

\begin{figure*}[!t]
  \centering
   \includegraphics[width=\linewidth]{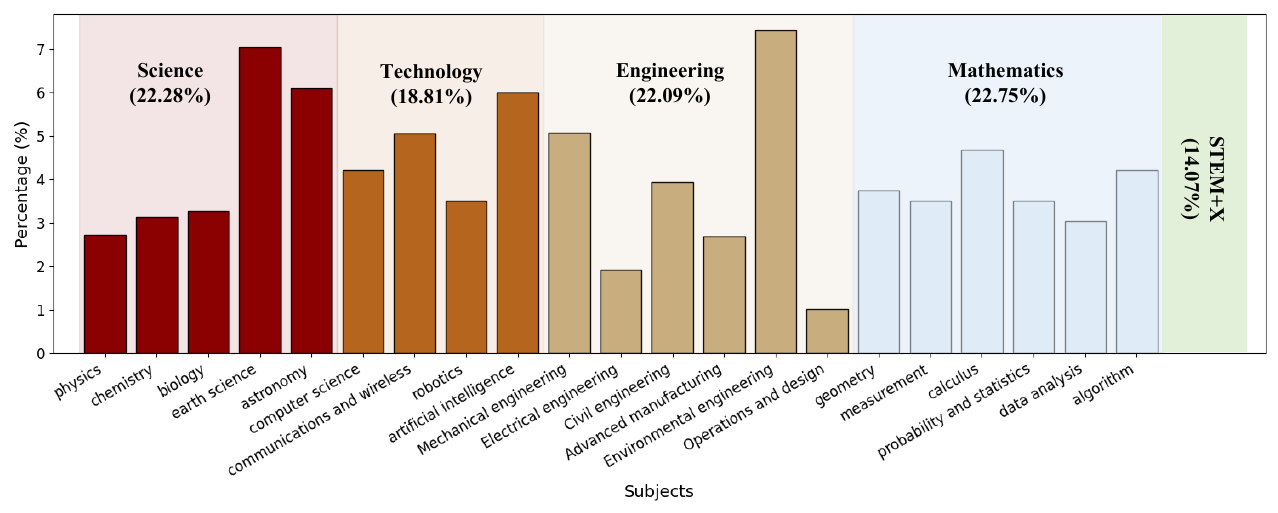}
   \vspace{-6mm}
   \caption{\textbf{Concept distribution of our SER dataset}. Our dataset exhibits a diverse distribution on different concept domains.}
   \label{fig:distribution}
   \vspace{-3mm}
\end{figure*}

\subsection{Query-based Retrieval}
Existing work in Query-based Image Retrieval (QBIR) primarily includes content-based image retrieval \cite{chen2022deep}, text-based image retrieval \cite{li2011text}, and multi-modal retrieval \cite{neculai2022probabilistic}. In content-based image retrieval, the visual features of images are directly utilized for retrieval. However, its reliance on fixed content and location makes it relatively inflexible in capturing diverse user intents \cite{lee-etal-2024-interactive}. Alternative methods like sketching \cite{chowdhury2022fs, chowdhury2023scenetrilogy} and scene graph construction \cite{johnson2015image} enable the retrieval of abstract images that are hard to describe verbally, though they lack the intuitive ease of natural language-based retrieval. In text-based image retrieval, enhancements to text queries often involve indicating content structure. However, these approaches are either restricted by closed vocabularies \cite{mai2017spatial, kilickaya2021structured} or face substantial challenges \cite{li2017generating} in deriving structures from natural language descriptions. Recent multi-modal approaches, such as cross-modal scene graph-based image-text retrieval \cite{wang2020cross} and joint visual-scene graph embedding for image retrieval \cite{belilovsky2017joint}, still depend on word embeddings and image features.

Despite advancements in QBIR, challenges including the semantic gap that can lead to inaccurate retrieval results, high computational complexity and resource costs for large-scale image databases, and the high cost of obtaining quality data annotations \cite{li-etal-2024-generative}. The application of QBIR to educational resource retrieval is promising but has been hindered by the complexity of educational discourse, the limitations of educational databases, and the associated costs \cite{zhou-etal-2024-vista}.  Our query model effectively combines multi-modal retrieval methods, integrating audio and natural language with multi-style image inputs. The former enables natural and rapid expression of content, while the latter facilitates accurate and intuitive image localization, enhancing educational data retrieval.

\subsection{Prompt Tuning}

Prompt tuning \cite{brown2020language} was first proposed in natural language processing~(NLP) and has been an efficient approach that bridges the gap between pre-trained language models and downstream tasks \cite{li2023weakly,zhao2023prompt,zhao2024unlearning}. Prompt tuning leverages natural language prompts to optimize the language model’s ability to understand tasks, which demonstrates exceptional performance in few-shot and zero-shot learning. 
Recent studies have focused on optimizing various components of prompt tuning, such as  prompt generation, continuous prompt optimization \cite{prompt3}, and adapting to large-scale models through methods like in-context learning \cite{dong2024survey,zhao2024universal}, instruction-tuning \cite{wang2024pandalm}, and chain-of-thought \cite{prompt4}. For example, \citet{lester2021power} leverage soft prompts to condition frozen language models to enhance the performance of specific downstream tasks. \citet{long2024prompt} propose an adversarial in-context learning algorithm, which leverages adversarial learning to optimize task-related prompts.

Furthermore, prompt tuning has gradually become a pivotal technique in computer vision \cite{shen2024multitask}, enabling efficient adaptation of pre-trained models to diverse tasks. Notable methods include visual prompt tuning for classification \cite{jia2022visual}, learning to prompt for continual learning \cite{wang2022learning}, context optimization and conditional prompt learning for multi-modal models \cite{zhou2022conditional}, and prompt-based domain adaptation strategies \cite{ge2023domain}.
For example, \citet{nie2023pro} introduce the pro-tuning algorithm for learning task-specific vision prompts, applied to downstream task input images with the pre-trained model remaining frozen.
\citet{shen2024multitask} leverage cross-task knowledge to optimize prompts, thereby enhancing the performance of vision-language models and avoiding the need to independently learn prompt vectors for each task from scratch.
\citet{cho2023distribution} introduce distribution-aware prompt tuning for vision-language models, optimizing prompts by balancing inter-class dispersion and intra-class similarity.
MaPLe \cite{khattak2023maple} further transfers text features to the visual encoder during prompt tuning to avoid overfitting. These approaches leverage learnable prompts to enhance model performance across various applications. Despite significant advancements in previous research, challenges remain in extracting semantic features from style-diversified images and optimizing templates within cont.inuous prompt tuning. 
In this study, we employ both NLP and visual prompt tuning to optimize STEM educational content retrieval, enhancing retrieval accuracy and efficiency by adjusting prompt tokens.

\begin{figure}[!tbp]
  \centering
   \includegraphics[width=\linewidth]{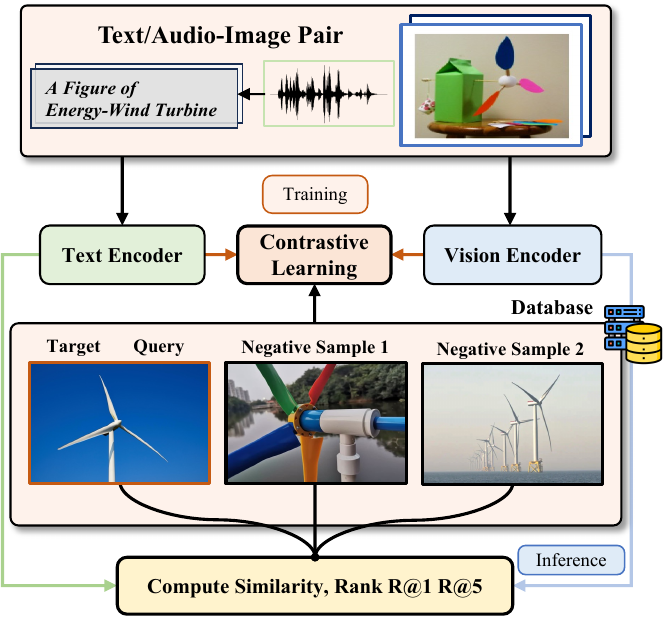}
   \vspace{-5mm}
   \caption{\textbf{The pipeline of Uni-Retrieval.} The image-text/audio pairs are input into their respective modality encoders. During the training procedure, contrastive learning is applied between the modality features of the positive samples~(image-text/audio pairs) and the negative samples. During the inference procedure, the model calculates the similarity between the modality features of the query and the embeddings stored in the database. The retrieved results are ranked, and the performance is evaluated using R@1/R@5 as metrics.}
   \label{fig:pipeline}
   \vspace{-5mm}
\end{figure}

\section{Motivation and Scenarios}
In practical teaching scenarios, teachers often encounter the need for precise image retrieval, such as searching for hand-drawn sketches, student-created artistic images, blurry blackboard drawings captured from a distance, classroom photographs of physical objects, or images from textbooks. However, current retrieval models predominantly focus on text-natural image queries, overlooking the diverse query styles common in educational contexts. This limitation makes it challenging for teachers to efficiently identify and retrieve educational images or texts tailored to diverse teaching scenarios, such as accurately setting learning contexts, articulating key teaching points, presenting instructional materials, and quickly locating supplementary resources.

Our proposed method enables teachers to query various styles' answers with a range of retrieval approaches, including text, image, audio, or combinations of these modalities. This approach ensures fast and convenient retrieval, significantly reducing preparation time for teaching. Once teachers input their queries, our Uni-Retrieval system employs contrastive learning to compare images and text, calculating similarities based on attributes like objects, shapes, quantities, and orientations. The system ranks all database entries by similarity, outputting the top-1 or top-5 results to identify the most relevant and accurate teaching resource images, as illustrated in Fig.~\ref{fig:pipeline}. This approach empowers teachers to manage complex and dynamic teaching scenarios effortlessly, enhancing the clarity and effectiveness of STEM education.

\section{Experiments}
\label{supsubsec:Experiment Settings}


In the database, the texts, their corresponding four images and audio structures for each dataset can share a single index, significantly reducing query time. All images and text are preprocessed using pretrained models to extract features, which are stored as embeddings. This approach eliminates the need for repeated feature extraction during use, saving time and reducing computational overhead, improving the efficiency of the retrieval system.

For the dataset selection, we choose four another datasets except our SER dataset, including the DSR dataset \cite{li2025freestyleret}, the ImageNet-X dataset, the SketchCOCO dataset \cite{gao2020sketchycoco} and the DomainNet dataset \cite{peng2019moment}. We use internVL-1.5 \cite{Chen_2024_CVPR} to annotate the paint/sketch caption for the SketchCOCO and the DomainNet dataset. For the model in the prototype learning module, we choose the VGG \cite{vgg} as the feature extractor. For the baseline selection, we apply two cross-modality pre-trained models (CLIP \cite{clip}, BLIP \cite{blip}), two multi-modality pre-trained models~(LanguageBind \cite{languagebind}, Unified-IO2 \cite{uio2}), two style retrieval models (SceneTrilogy \cite{scenetrilogy}, FashionNTM \cite{fashionntm}), four most recent cross-modality prompt learning models (VPT \cite{jia2022visual}, CoCoOP \cite{zhou2022conditional}, MaPLe \cite{khattak2023maple}, FreestyleRet \cite{li2025freestyleret}), and two database-driven retrieval models (GASKN \cite{GASKN}, MKG \cite{mkg}) for the fair comparison. Specifically, we fine-tune the cross-modality models (CLIP, BLIP) on SER for convergence. We also train the prompt learning models on SER dataset based on VPT's settings to adapt STEM style-diversified inputs. As for the multi-modality models, we evaluate the zero-shot performance on the style-diversified STEM education retrieval task due to multi-modality models' comprehensionability on multi-style image inputs.

For the experiments on the SER dataset, Uni-Retrieval is initialized with OpenCLIP's weights and trained on 8 A100 GPUs with batch size 24 per GPU and 20 training epochs. We use AdamW as the optimizer, set the learning rate to 1e-5 with a linearly warmed up operation in the first epochs and then decayed by the cosine learning rate schedule. The seed is set as 42. What's more, all input images are resized into $224\times224$ resolution and then augmented by normalized operation. All text are padding zero to the max length of 20.

For the fine-tuning CLIP and BLIP models, all experiment settings are the same as Uni-Retrieval except the learning rate is set as 1e-6. For prompt tuning models, we both use 4 prompt tokens to expand the token sequence. For all transformer-based models, we use the ViT-Large and 24-layers text transformer as the foundation models to keep balance between performance and efficiency.

\end{document}